# Electronic properties of low dimensional structures

**Azzedine Bendounan**

*Experimentelle Physik II, Universität Würzburg, Am Hubland, D-97074 Würzburg, Germany,*

*Synchrotron Soleil, St. Aubin, BP48, F-91192 Gif-sur-Yvette Cedex, France*

azzedine.bendounan@synchrotron-soleil.fr

**Abstract:**
**Exotic phenomena about the behavior of electrons inside the solid were a long time ago predicted by the quantum mechanic physics and are only recently experimentally observed, in particular for systems of extremely reduced dimensions. Here, I report on recent experimental observation of fundamental effect concerning the dispersion properties of the surface state influenced by the presence of surface reconstruction.**

Many physical properties of the materials, concerning for example the transport, are directly connected to theirs electronic band structures and are mainly controlled by the behavior of the electronic states located in the vicinity of the Fermi level. Perhaps one can refer to the *superconductivity* as the most prestigious phenomenon thoroughly investigated over the last decades in order to draw a clear picture of the underlying physics. Much interest has also been devoted to the behavior of the *surface states*, which were successfully used as a model system to explore different electronic interactions in the solid (electron-electron, electron-phonon and electron-defect) [1] and have also recently open up the way for new spintronic developments [2]. Surface state represents a peculiar solution of Schrödinger equation with energy in the gap of the bulk band structure and imaginary momentum of the wave function perpendicularly to the surface. Localized in few topmost layers near the surface, it is completely decoupled from the bulk states and has a Bloch wave character with nearly-free-electron like dispersion parallel to the surface. Such paradigmatic states have been studied spectroscopically by angle resolved photoemission (ARPES) [3] and have also been visualized formant standing wave patterns in scanning tunneling microscopy (STM) [4].

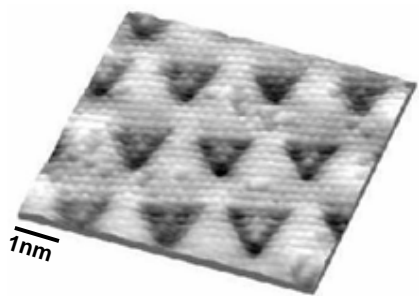

*Fig.01: STM image of Cu(111) surface covered by one Ag monolayer. The surface presents triangular-like structure, which develops due to a large lattice mismatch between Ag and Cu [5].*

The here presented data have been obtained on one monolayer of silver (Ag) deposited under high vacuum conditions on a (111)-oriented copper substrate, i.e. Cu(111). The silver atoms form epitaxially and ordered mono-atomic layer characterized by a triangular like superstructure with (9×9) super-periodicity, which results from the large lattice mismatch between Ag and Cu (13%) [5]. Obtained by STM technique, an image of the surface — shown in Fig.01 — displays the atomic arrangement of the Ag over-layer. One observes uniform distribution of triangles with hexagonal symmetry resulting from a lack of some Ag-atoms developing through the relaxation of the surface strain.
The surface band structure is measured by ARPES. In Fig.02 the intensity of the emitted electrons is presented as function of their binding energy and their wave vector. The dispersion is measured along the $\overline{\Gamma} - \overline{M}$ direction of the surface Brillouin zone (SBZ). We see a band with parabolic dispersion characterized by an abrupt drop of the photoemission intensity at $k \approx \pm\ 0.15$ Å$^{-1}$, which restores close to Fermi level ($E_F$) [6]. This intensity missing is interpreted as an opening of a band gap at the boundaries of the reduced-SBZ originated from the translational symmetry of the (9×9) surface reconstruction. In fact, the electrons undergo a Bragg like diffraction at the gap position.

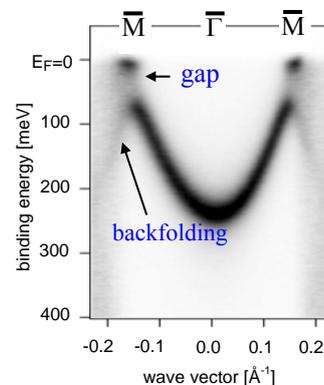

*Fig.02: Surface state band measured by photoemission on Cu(111) surface covered by one Ag monolayer. The image shows the intensity of the photoelectrons as function of the binding energy and the wave vector* [6].

A back-folding of the surface state occurs around the high-symmetry points of the reduced SBZ. The back-folded intensity depends on the strength of the superstructure potential and appears high only near the intersection at the SBZ boundaries [7]. This effect is not observed on non-reconstructed surface where isotropic band dispersion was usually obtained. It represents a direct experimental observation of fundamental phenomenon concerning the behavior of the electronic states inside the solid. The dispersion of the surface state follows perfectly the principles of nearly free electrons model, which is well described in the text books of solid state physics [8].

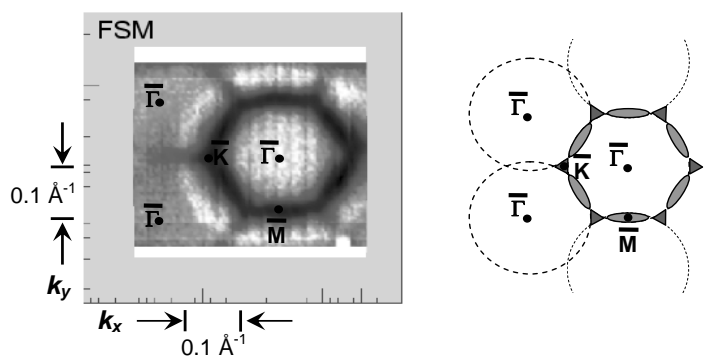

*Fig.03: left panel: Fermi surface map (FSM) measured by ARPES on 1ML Ag/Cu(111) showing a sixfold symmetry resulting from an overlap of original and the adjacent FSMs, as it is sketched in the right panel* [6].

The Fermi surface map (FSM) — presented in Fig.03, left panel — is obtained by mapping of the photoemission intensity at $E_F$ over the reduced-SBZ. A sixfold symmetry of the intensity distribution is observed, whereas on clean Cu(111), the FSM looks like a nearly perfect circle in the corresponding wave vector range [3]. In fact, the reduced size of SBZ leads near the zone boundaries to an overlap of the originally and the adjacent circular Fermi surfaces.